\documentclass{pas}

\usepackage{layout}

\usepackage{xcolor} 
\usepackage{hyperref}
\hypersetup{
    colorlinks=true,
    linkcolor=blue,
    citecolor=violet,
    urlcolor=teal
}
\usepackage{longtable}
\usepackage{amsmath}
\usepackage{cprotect}

\usepackage{graphicx}
\usepackage{overpic}
\usepackage{subcaption}
\usepackage{amssymb,microtype,siunitx,booktabs}
\usepackage{multirow}
\usepackage[T1]{fontenc}
\usepackage{ae,aecompl}

\usepackage[separate-uncertainty=true,multi-part-units=single]{siunitx}
\usepackage{threeparttable}
\usepackage{outlines}
\usepackage{lastpage}

\usepackage{aas-macros}

\crefname{figure}{Figure}{Figures}
\crefname{table}{Table}{Tables}
\crefname{equation}{Equation}{Equations}

\newcommand\rah{\mbox{$^{\mathrm h}$}}%
\newcommand\ram{\mbox{$^{\mathrm m}$}}%
\newcommand\ras{\mbox{$^{\mathrm s}$}}%

\newcommand\arcdeg{\mbox{$^\circ$}}%
\newcommand\arcmin{\mbox{$^\prime$}}%
\newcommand\arcsec{\mbox{$^{\prime\prime}$}}%

\DeclareSIUnit\jansky{Jy}
\DeclareSIUnit\beam{beam}
\DeclareSIUnit{\mjpb}{\milli\jansky\per\beam}

\newcommand{\nodata}{--}

\newcommand{\changed}[1]{#1}

\newcommand{\WWPsA}{WW\;PsA}
\newcommand{\TXPsA}{TX\;PsA}
\newcommand{\UVCet}{UV\;Cet}

\newcommand{\CUVir}{CU\;Vir}
\newcommand{\ADLeo}{AD\;Leo}

\newcommand{\ion}[2]{#1\,\textsc{#2}}
\newcommand{\HI}{\ion{H}{i}}
\newcommand{\Mdwarfs}{M~dwarfs}

{%
  \end{outline}
}

\DeclareSIUnit\jansky{Jy}
\DeclareSIUnit\mas{mas}
\DeclareSIUnit\year{yr}
\DeclareSIUnit\hour{h}
\DeclareSIUnit\minute{min}
\DeclareSIUnit\deg{deg}
\DeclareSIUnit\au{AU}
\DeclareSIUnit\parsec{pc}
\DeclareSIUnit\erg{erg}
\DeclareSIUnit\gauss{G}
\DeclareSIUnit\percent{\%}
\DeclareSIUnit\beam{beam}
\DeclareSIUnit\angstrom{\text {Å}}
\DeclareSIUnit{\radlum}{\erg\per\second\per\hertz}
\DeclareSIUnit{\mjpb}{\milli\jansky\per\beam}
\DeclareSIUnit{\Rsun}{R_\odot}
\DeclareSIUnit{\Rjup}{R_{\rm J}}
\DeclareSIUnit{\Mjup}{M_{\rm J}}

\def\tablefont{\fontfamily{\sfdefault}\fontsize{7.5}{9}\selectfont}%

\begin{document}

\lefttitle{Publications of the Astronomical Society of Australia}
\righttitle{J.\ Mitchell-Bolton \textit{et al.}}

\jnlPage{1}{\pageref*{LastPage}}
\jnlDoiYr{2026}
\doival{10.1017/pasa.xxxx.xx}

\articletitt{Research Paper}

\newcommand{\papertitle}{Coherent radio bursts, polarity reversal, and quiescent emission from the M4V~dwarf \WWPsA}
\title{\papertitle}

\author{\sn{Jackson} \gn{Mitchell-Bolton}$^{1}$, 
\sn{Laura} \gn{Driessen}$^{1}$,
\sn{Tara} \gn{Murphy}$^{1,2}$,  
\sn{Joshua} \gn{Pritchard}$^{3}$,
\sn{Barnali} \gn{Das}$^{4}$,
\sn{Richard} \gn{Dodson}$^{5}$,
\sn{Martin} \gn{Meyer}$^{5}$,
\sn{Jonghwan} \gn{Rhee}$^{6,5}$ and 
\sn{Yuanming} \gn{Wang}$^{7,2}$
}

\affil{
$^1$Sydney Institute for Astronomy, School of Physics, The University of Sydney, NSW 2006, Australia, 
$^2$ARC Centre of Excellence for Gravitational Wave Discovery (OzGrav), Australia, 
$^3$CSIRO Space and Astronomy, PO Box 76, Epping, NSW 1710, Australia,
$^4$National Centre for Radio Astrophysics, Tata Institute of Fundamental Research, Pune University Campus, Pune-411007, India,
$^5$International Centre for Radio Astronomy Research (ICRAR), University of Western Australia, 35 Stirling Hwy, Crawley, WA 6009, Australia,
$^6$Australia Telescope National Facility, CSIRO Space \& Astronomy, P.O.\ Box 1130, Bentley, WA 6102, Australia and
$^7$Centre for Astrophysics and Supercomputing, Swinburne University of Technology, Hawthorn, VIC 3122
}

\corresp{Jackson Mitchell-Bolton, Email: jmit6654@uni.sydney.edu.au}

\citeauth{Mitchell-Bolton J, Driessen L, Murphy T, Pritchard J, Das B, Dodson R, Meyer M, Rhee J, Wang Y,
\papertitle. 
{\it Publications of the Astronomical Society of Australia} {\bf 00}, 1--\pageref*{LastPage}. https://doi.org/10.1017/pasa.xxxx.xx}

\history{(Received xx xx xxxx; revised xx xx xxxx; accepted xx xx xxxx)}

\begin{abstract}
Coherent radio emission from magnetically active \Mdwarfs\ is largely observed in two varieties: stochastic solar-like bursts and beamed, rotationally modulated emission including auroral pulses. Detailed monitoring of these stars is critical for probing the plasmas and magnetic fields driving the emission.
We conducted a search for polarised radio emission from stars within \SI{85}{\parsec}, covering \SI{2.2e4}{\hour} of stellar dynamic spectra from the ASKAP Deep Investigation of Neutral Gas Origins (DINGO) survey.
We report the discovery of coherent radio bursts from the M4V dwarf \WWPsA. In addition to quiescent emission, we detect three long-duration ($\gtrsim\text{\SI{30}{\minute}}$) bursts, each $\sim$100\% circularly polarised, in \SI{283.3}{\hour} of data. \changed{We also report a fourth burst in an epoch not covered by our search, detected using the VASTER pipeline. One of the} bursts features a polarity reversal midway through, \changed{based on which we favour the electron cyclotron maser (ECM) mechanism over plasma emission}. We \changed{tentatively propose the bursts are due to} rotationally modulated ECM pulses of variable intensity. We infer \changed{the required} physical properties of \WWPsA, including a high obliquity $\beta = \text{74}^{+\text{6}}_{-\text{22}}\,\text{\SI{}{\deg}}$ under an oblique rotator model.
\end{abstract}


\maketitle

\newpage

\section{Introduction}

At radio frequencies, magnetically active \Mdwarfs\ are capable of emission with intensity far exceeding that from the Sun. In addition to slowly-varying quiescent emission ascribed to gyrosynchrotron radiation \citep[e.g.][]{Kundu&Shevgaonkar1985,Pallavicini1985,Williams2014}, many \Mdwarfs\ produce transient bursts caused by a coherent mechanism: electron cyclotron maser (ECM) emission \citep[e.g.][]{Gary1982,Kundu1987,Villadsen&Hallinan2019} or plasma emission \citep[e.g.][]{Stepanov2001,Osten&Bastian2006,Callingham2025}. 

A central question in the study of this emission regards the transition along the main sequence from solar-like bursts, primarily originating from plasma emission, to the rotationally modulated ECM emission exhibited by ultracool dwarfs \citep[e.g.][]{Hallinan2007,Lynch2015,Hallinan2015,Rose2023}. Both have been observed from \Mdwarfs\ \citep[e.g.][]{Villadsen&Hallinan2019,Zic2019,Zic2020,Callingham2025}. 
Detailed monitoring of individual \Mdwarfs\ is instrumental for identifying each type, and can be used to probe the plasmas and magnetic fields from which the emission originates \citep[e.g.][]{Villadsen&Hallinan2019,Bloot2024,Callingham2025}. 
This is of particular interest for dwarfs of spectral type $\sim$M4 \citep[e.g.][]{Morin2008}, with masses near the $\sim0.35\,M_\odot$ transition in interior structure below which stars are fully convective \citep{Chabrier&Baraffe1997}.

Coherent, rotationally modulated emission from \Mdwarfs\ can be auroral in nature \citep{Hallinan2015,Zic2019}, corresponding to a dipolar magnetic field \citep[see the model presented by][]{Leto2016}, but in some cases is better modelled with more complicated magnetic field configurations \citep{Lynch2015}. Further complicating the matter, the stochastic bursts from the M1Ve~dwarf AU~Mic were recently found to have variations in fractional polarisation matching the rotation period, explained by auroral ECM emission \citep{Bloot2024}.

In recent years, the Australian SKA Pathfinder \citep[ASKAP;][]{Hotan2021} has enabled the discovery of many radio stars \citep[e.g.][]{Pritchard2021,Pritchard2024,Rose2023,Driessen2024,Das2025}. Many of the ASKAP surveys used in these radio star searches have integration times of 12--\SI{15}{\minute} \citep[e.g.][]{McConnell2020,Murphy2021}, allowing rapid coverage of wide areas of the sky. However, for detailed investigations of individual stars, such surveys are limited in the amount of observation time they contribute.

In contrast, the ASKAP Deep Investigation of Neutral Gas Origins survey \citep[DINGO;][]{Rhee2023} targets two adjacent $\sim$\SI{30}{deg$^2$} tiles of extragalactic sky, with a planned 1600~hours of observation time on each tile, split between two frequency ranges: $\text{Low-}z$ targeting the \SI{1420}{MHz} \HI\ line at redshifts $0 < z < 0.1$, and $\text{High-}z$ targeting $0.25 < z < 0.4$.
This presents an opportunity for long-duration radio monitoring of any radio stars present in the DINGO survey area. 

In this paper we present a detailed investigation of the radio emission from the M4V~dwarf \WWPsA\, including coherent bursts detected in a search for radio emission from stars within \SI{85}{\parsec} with DINGO. In \cref{sec:obs} we detail our observations and data analysis, including our search method as well as our analysis of the bursts and quiescent emission from \WWPsA. In \cref{sec:discussion} we discuss the significance of our search for the population statistics of nearby radio stars, describe the known properties of \WWPsA, and interpret the observed emission. In particular, we investigate the possibility that the bursts are due to rotationally modulated ECM emission.

\section{Observations and data analysis}\label{sec:obs}

\subsection{Search method and detection of \WWPsA}\label{sec:search}

\begin{figure*}
    \centering
    \includegraphics[width=\textwidth]{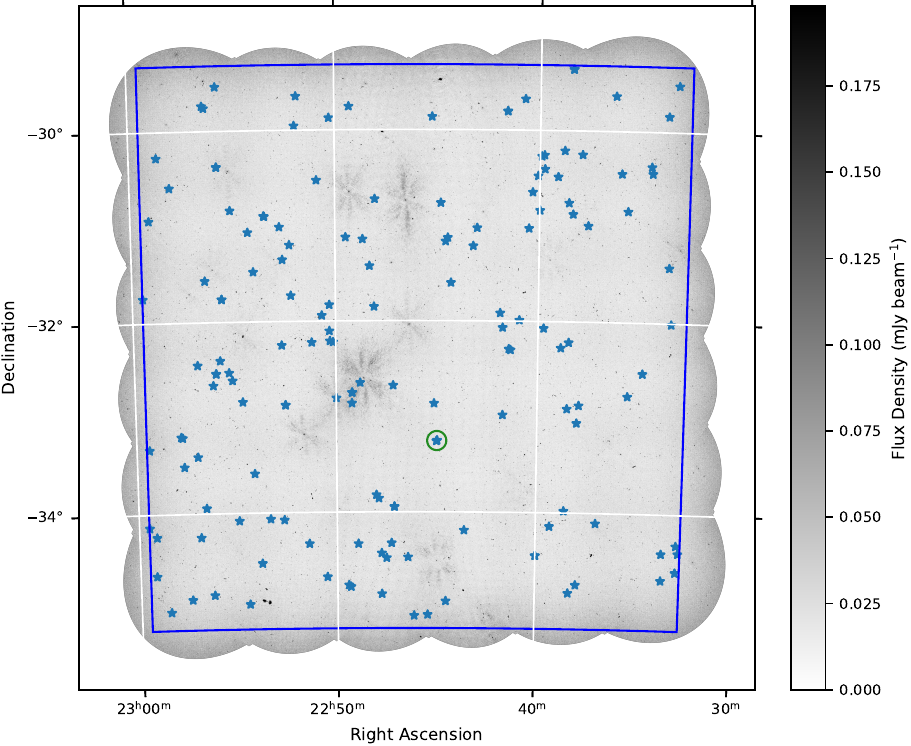}
    \caption{Positions of all 154 GCNS stars with median distance estimates \citep{GaiaCollaboration2021GCNS} below \SI{85}{pc}, shown as blue stars, selected within the bounds marked by the dark blue rectangle for the search described in \cref{sec:search}. \WWPsA\ and its binary companion \TXPsA\ are marked by the green circle. \textit{Background:} Stokes~$I$ image showing Footprint~A of DINGO Tile~0 \citep[see][]{Rhee2023} from the ASKAP scheduling block SB65210. The positions of the stars are proper motion corrected to the epoch J2024.7, corresponding to this observation.}
    \label{fig:dingo_GCNS_StarCoords}
\end{figure*}

To construct a sample of stars for our search, we selected all objects in the \textit{Gaia} Catalogue of Nearby Stars~\citep[GCNS;][]{GaiaCollaboration2021GCNS} within a distance of \SI{85}{pc} in the \SI{33.45}{\square\deg} region defined by the bounds $22\rah 32\ram 40\ras < \alpha < 22\rah 59\ram 30\ras$ and $-35\arcdeg 13\arcmin < \delta < -29\arcdeg 19\arcmin$. This region corresponds to Footprint\;A of DINGO Tile\;0 \citep[see][]{Rhee2023}, as shown in \cref{fig:dingo_GCNS_StarCoords}. We used the J2016.0 coordinates and median distance estimates \verb"dist_50" provided in the GCNS \citep{GaiaCollaboration2021,GaiaCollaboration2021GCNS} to select the sample. This resulted in 154 stars, the positions of which are also shown in \cref{fig:dingo_GCNS_StarCoords}. 

For each star, we used \textsc{DStools}\footnote{\href{https://github.com/askap-vast/dstools}{https://github.com/askap-vast/dstools}} \citep{DStoolsv2.0.0} to produce a dynamic spectrum and lightcurve in Stokes~$I$, $Q$, $U$, and~$V$\footnote{Throughout this paper, we adopt the IAU/IEEE definition of Stokes~$V$.} at the proper motion corrected coordinates in each of the 18 DINGO full-survey observations (project code AS204) listed in \cref{tab:WWPsAobservations}. Although our sample of stars was selected based on Footprint\;A of DINGO Tile\;0, some of these observations were conducted using Footprint\;B. The offset between the footprints was small enough to allow dynamic spectrum creation for all stars in our sample in each epoch. This amounted to \SI{144}{\hour} of data per object. 

We produced the dynamic spectra and lightcurves using both the original \SI{10}{\second} time resolution and a time resolution of \SI{5}{\minute}, in both cases using the original \SI{1}{\mega\hertz} frequency resolution of the data. We flagged some bands with persistent radio frequency interference (RFI) using \textsc{DStools} (from the bottom of the observed frequency range to \SI{1315}{MHz} and the band 1379--\SI{1382}{MHz}), 
in addition to the flagging present in the visibility data as obtained from the CSIRO ASKAP Science Data Archive\footnote{\href{https://research.csiro.au/casda/}{https://research.csiro.au/casda/}} (CASDA). We performed a visual inspection of the Stokes~$V$ dynamic spectra and lightcurves, not including the other Stokes parameters as they contained artefacts caused by sidelobes from bright neighbouring sources (especially in Stokes~$I$). 
We did not find any emission above a $5\sigma$ detection threshold except for two bursts from \WWPsA\ (see \cref{sec:WW_PsA_physical_properties} for a description of this star). 

\begin{figure*}
    \centering
    \includegraphics[width=\textwidth]{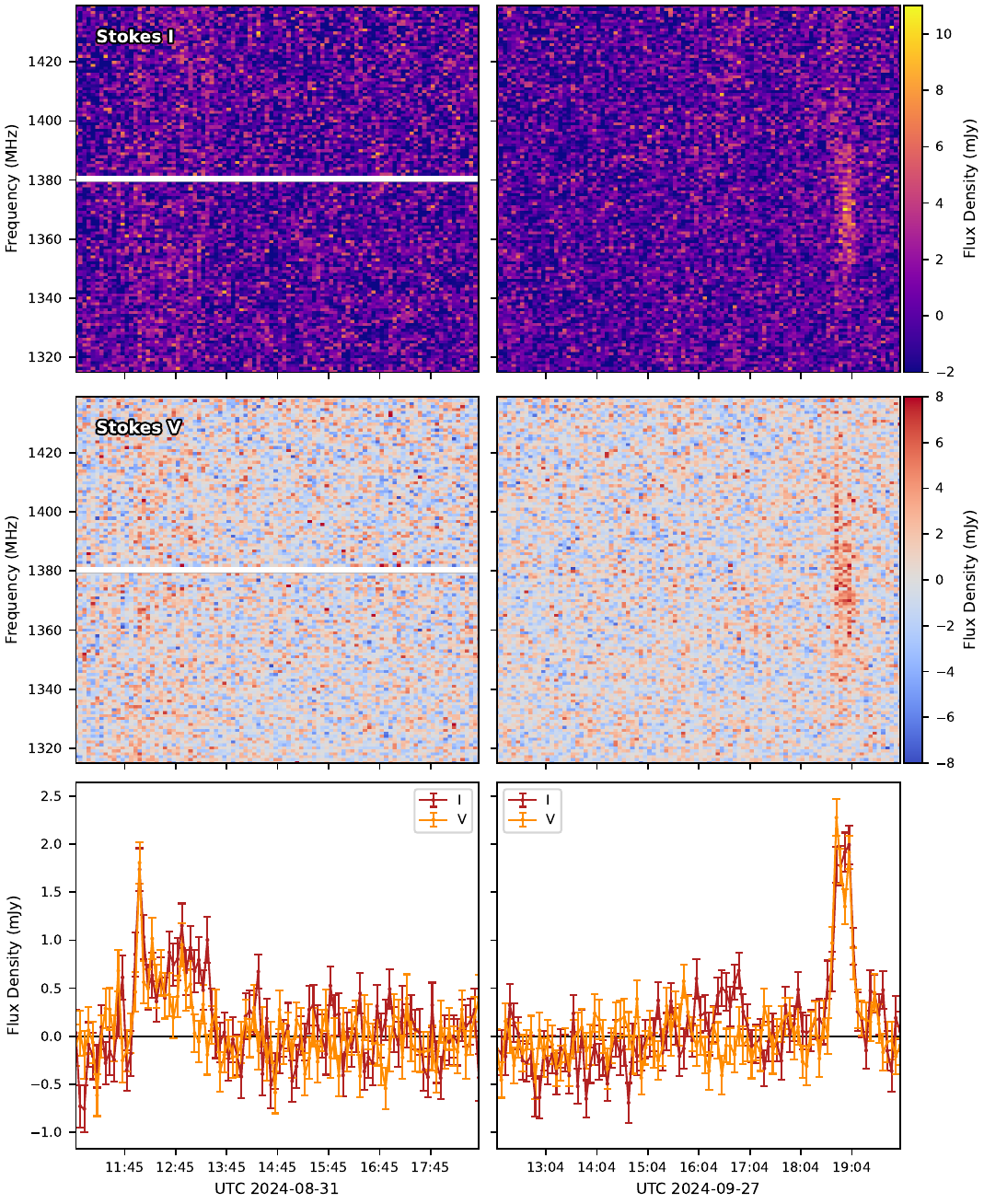}
    \caption{\textit{Top \& Middle}: Stokes~$I$ and $V$ dynamic spectra of \WWPsA\ in scheduling blocks SB65405 (left) and SB66106 (right) after field modelling and subtraction, with a time resolution of \SI{5}{\minute}. Visibilities from baselines shorter than \SI{500}{\metre} have been excluded. \textit{Bottom}: Stokes~$I$ and $V$ lightcurves for the same observations.}
    \label{fig:WW_PsA_I_V_ds_and_lc_SB65405_SB66106_tav30}
\end{figure*}

We used \textsc{DStools} to perform field modelling and subtraction for the two epochs (scheduling blocks SB65405 and SB66106) in which we found bursts from \WWPsA. This greatly reduced the impact of sidelobes from bright nearby sources, allowing us to make use of the Stokes~$I$, $Q$, and $U$ data for these bursts. We did not perform this step in the original search as it is computationally intensive. 
The resulting dynamic spectra and lightcurves in Stokes~$I$ and $V$ are shown in \cref{fig:WW_PsA_I_V_ds_and_lc_SB65405_SB66106_tav30}. The burst in SB65405 had a longer duration than that in SB66106, at $\sim$\SI{1.5}{\hour} and $\sim$\SI{0.5}{\hour}, respectively. The latter was much more strongly confined in frequency, being brightest in the range 1350--\SI{1410}{MHz}, with the former showing little variation in flux density across the observed band. Both had (signed) fractional circular polarisation $S_V/S_I \approx 1$, however this decreased to $S_V/S_I \approx 0$ towards the end of the burst in SB65405. We did not detect linear polarisation in either case.

\begin{table}
  \centering
  \caption[Summary of radio observations of \WWPsA]{
    Summary of radio observations searched for polarised emission from \WWPsA\ by producing dynamic spectra and lightcurves. Observations under the project code AS204 were included in the search for emission from nearby stars described in \cref{sec:search}.
  }\label{tab:WWPsAobservations}
    {\tablefont\begin{tabular}{@{\extracolsep{\fill}}lccccc}
      \toprule
      UT Start Time & Project & SBID & $\nu_{\rm obs}$ & $t_{\rm obs}$ \vspace{-4pt} & Notes \\
             & Code & & (\si{\mega\hertz}) & (\si{\hour}) & \\
      \hline
        2019--03--08 00:29:28 & AS034 & SB8132 & 744.5--1032.5 & 9.0 & \\
        
        2021--09--18 09:50:34 & AS104 & SB32214 & 1296.5--1440.5 & 8.0 \\
        2021--09--25 11:31:06 & AS104 & SB32511 & 891.5--1147.5 & 7.9 \\
        2021--12--05 07:31:54 & AS104 & SB34175 & 891.5--1147.5 & 7.9 \\
        2022--04--20 19:52:04 & AS104 & SB39697 & 1296.5--1440.5 & 8.0 \\
        2022--04--21 19:51:39 & AS104 & SB39722 & 1296.5--1440.5 & 8.0 \\
        2022--04--28 19:35:56 & AS104 & SB39995 & 1296.5--1440.5 & 8.0 \\
        2022--04--29 18:54:42 & AS104 & SB40018 & 1296.5--1440.5 & 8.0 \\
        2022--05--03 18:59:46 & AS104 & SB40196 & 1296.5--1440.5 & 8.0 \\
        2022--05--07 18:37:45 & AS104 & SB40329 & 891.5--1147.5 & 7.9 \\
        2022--05--10 19:45:55 & AS104 & SB40479 & 891.5--1147.5 & 7.9 \\
        2022--05--13 18:34:14 & AS104 & SB40607 & 891.5--1147.5 & 7.9 \\
        2022--05--17 20:15:51 & AS104 & SB40712 & 891.5--1147.5 & 7.9 \\
        2022--05--18 18:14:51 & AS104 & SB40733 & 891.5--1147.5 & 7.9 \\
        2022--05--19 18:13:18 & AS104 & SB40765 & 891.5--1147.5 & 7.9 \\
        2022--05--21 17:51:56 & AS104 & SB40848 & 891.5--1147.5 & 7.9 & a,b\\
        2022--05--22 19:48:04 & AS104 & SB40880 & 891.5--1147.5 & 7.9 \\
          
        2024--08--26 10:59:27 & AS204 & SB65177 & 1296.5--1440.5 & 8.0 & \\
        2024--08--27 10:52:36 & AS204 & SB65210 & 1296.5--1440.5 & 8.0 & \\
        2024--08--28 10:51:11 & AS204 & SB65268 & 1296.5--1440.5 & 8.0 & \\
        2024--08--29 11:20:01 & AS204 & SB65297 & 1296.5--1440.5 & 8.0 & \\
        2024--08--30 10:50:31 & AS204 & SB65347 & 1296.5--1440.5 & 8.0 & \\
        2024--08--31 10:44:26 & AS204 & SB65405 & 1296.5--1440.5 & 8.0 & a,b \\
        2024--09--01 12:25:57 & AS204 & SB65453 & 1296.5--1440.5 & 8.0 & \\
        2024--09--06 10:46:14 & AS204 & SB65665 & 1296.5--1440.5 & 8.0 & \\
        2024--09--07 13:45:08 & AS204 & SB65681 & 1296.5--1440.5 & 8.0 & \\
        2024--09--08 10:23:04 & AS204 & SB65690 & 1296.5--1440.5 & 8.0 & \\
        2024--09--12 09:48:08 & AS204 & SB65765 & 1296.5--1440.5 & 8.0 & \\
        2024--09--13 11:19:29 & AS204 & SB65818 & 1296.5--1440.5 & 8.0 & \\
        2024--09--14 09:48:59 & AS204 & SB65836 & 1296.5--1440.5 & 8.0 & \\
        2024--09--15 09:59:43 & AS204 & SB65853 & 1296.5--1440.5 & 8.0 & \\
        2024--09--24 11:37:26 & AS204 & SB66032 & 1296.5--1440.5 & 8.0 & \\
        2024--09--25 11:28:08 & AS204 & SB66053 & 1296.5--1440.5 & 8.0 & \\
        2024--09--27 12:04:09 & AS204 & SB66106 & 1296.5--1440.5 & 8.0 & a,b \\
        2024--09--28 12:02:35 & AS204 & SB66134 & 1296.5--1440.5 & 8.0 & \\

        2026--03--15 00:42:52 & C3363 & \nodata & 1152--3072 & 3.3 & a
      \botrule
    \end{tabular}}
    \begin{tabnote}
    {(a) Stokes I dynamic spectrum and lightcurve produced and inspected.}\tnp
    {(b) Circularly polarised burst detected.}\tnp
    \end{tabnote}
\end{table} 

\subsection{Additional archival observations of \WWPsA}\label{sec:archival_observations}

We searched for circularly polarised emission from \WWPsA\ in one observation from the ASKAP Early Science Broadband Survey \citep{Harvey-Smith2018,O'Brien2019} and 16 from the DINGO pilot survey (project codes AS034 and AS104, respectively, in \cref{tab:WWPsAobservations}), totalling 137~hours. Again, we performed this search by visual inspection of the dynamic spectra and lightcurves in Stokes~$V$. We detected one additional event: a burst in SB40848 lasting $\sim$\SI{2}{hr}\;\SI{35}{\minute} with a reversal in the sense of its circular polarisation halfway through, from right-handed to left-handed. Prior to this polarity reversal, this burst reached a maximum Stokes~$V$ flux density $S_V = 5.9 \pm 0.2$\,mJy, significantly brighter than those in SB65405 and SB66106. After the reversal, the burst was much fainter, reaching $S_V = -1.6 \pm 0.2$\,mJy, comparable in magnitude to the bursts in \cref{fig:WW_PsA_I_V_ds_and_lc_SB65405_SB66106_tav30}. We did not detect linear polarisation.

\begin{figure*}
    \centering
    \includegraphics[width=\textwidth]{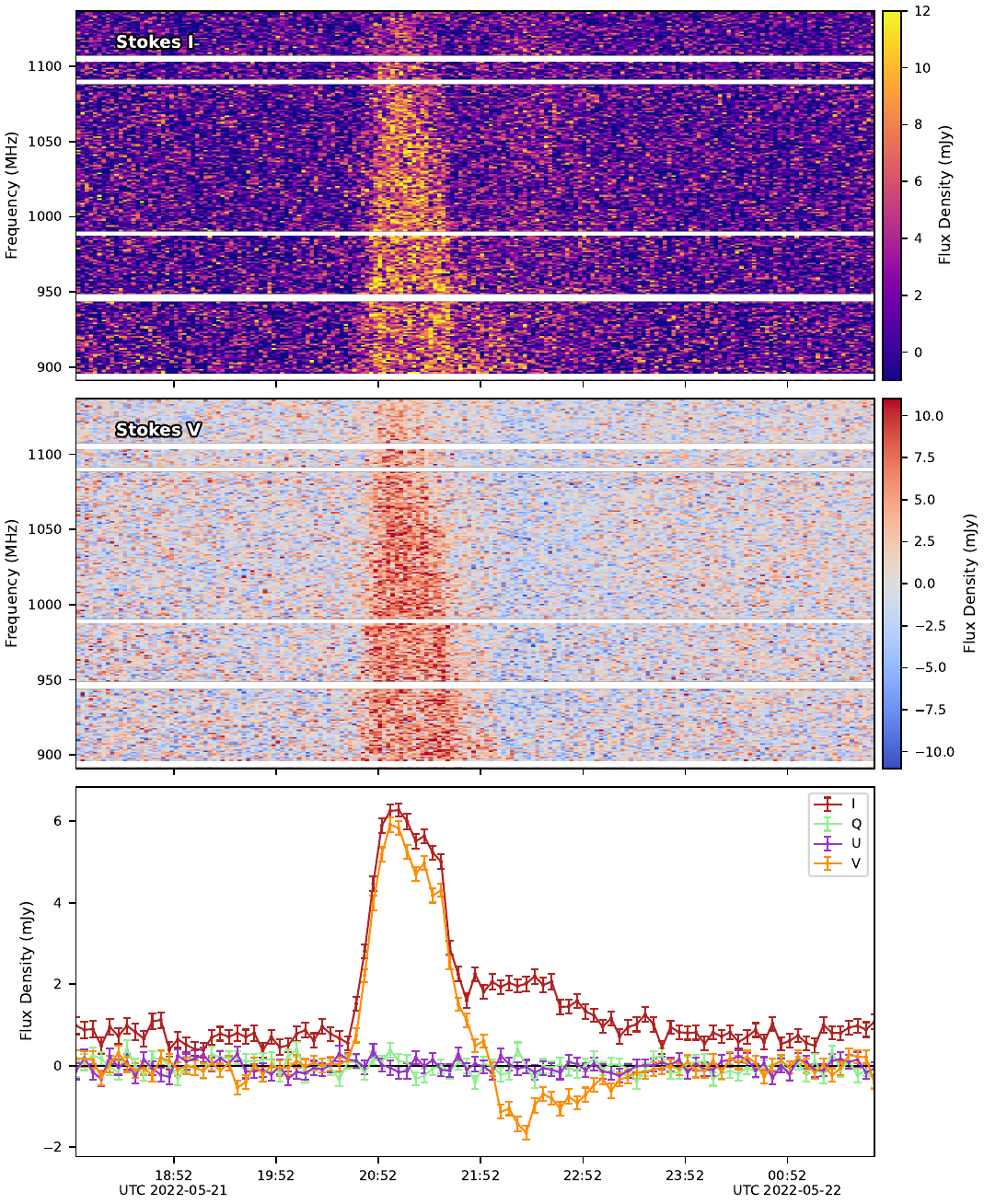}
    \caption{\textit{Top \& Middle}: Stokes~$I$ and $V$ dynamic spectra of \WWPsA\ in scheduling block SB40848 after field modelling and subtraction, with a time resolution of \SI{2.5}{\minute}. Visibilities from baselines shorter than \SI{500}{\metre} have been excluded. \textit{Bottom}: Stokes~$I$, $Q$, $U$, and $V$ lightcurves for the same observation with a time resolution of \SI{5}{\minute}. 
    }
    \label{fig:WW_PsA_flip}
\end{figure*}

The Stokes~$I$ and $V$ dynamic spectra and lightcurves from SB40848, after field modelling and subtraction (and excluding visibilities from the baselines shorter than \SI{500}{m}), are shown in \cref{fig:WW_PsA_flip}. The Stokes~$I$ lightcurve reveals quiescent emission before and after the burst. We discuss the emission seen in this observation further in \cref{sec:SB40848_discussion}.

This search, as well as the search described in \cref{sec:search}, only included observations from 2024 and earlier. However, we subsequently detected a burst from the later DINGO observation SB76830 using the VASTER pipeline \citep{Wang2026}. This observation covered the band 1296.5--\SI{1440.5}{\mega\hertz}. The burst began at 2025--09--13 $\sim$16:30~UTC and continued until the end of the observation at $\sim$21:15~UTC, making it $\gtrsim\text{\SI{4}{\hour}\,\SI{45}{\minute}}$ in duration. Its peak Stokes~$I$ flux density in the image plane, obtained from VASTER, reached a maximum of \SI{3.4(2)}{\mjpb} at $\sim$19:15~UTC. We leave a detailed analysis of this burst for future work, but discuss the implications of its time of arrival in \cref{sec:rotational_modulation}.

\subsection{Follow-up observation}\label{sec:follow_up}

\begin{figure}
    \centering
    \includegraphics[width=\linewidth]{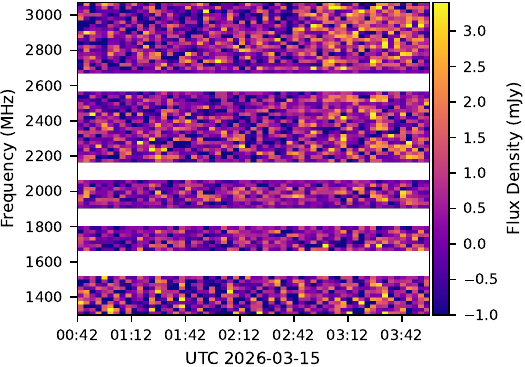}
    \caption{Stokes~$I$ dynamic spectrum from our follow-up observation of \WWPsA\ with the ATCA. The flux density of the quiescent emission is greater at higher frequencies and increases towards the end of the observation.}
    \label{fig:ATCA_qui}
\end{figure}

We conducted a 3.5-hour radio follow-up observation of \WWPsA\ with the Australia Telescope Compact Array (ATCA) on 2026-03-15 (project code C3363). This observation was performed in the 6C configuration using the \SI{16}{cm} (L-band) receiver with a \SI{1920}{MHz}
band centred on \SI{2112}{MHz}. We used PKS~B1934$-$638 (PKS~J1939$-$6342) to calibrate the bandpass response, flux scale, and polarisation leakage. We performed phase calibration with interleaved scans of PKS~B2245$-$328 (PKS~J2248$-$3235), each lasting \SI{90}{\second}. We flagged and calibrated the data using standard continuum data processing routines in \textsc{miriad}\footnote{\href{https://github.com/csiro/miriad}{https://github.com/csiro/miriad}} \citep{Sault1995} and produced Stokes~$I$ and $V$ dynamic spectra. We detected quiescent emission in Stokes~$I$, as shown in \cref{fig:ATCA_qui}, but we did not make a detection in Stokes~$V$. The quiescent emission was brighter at higher frequencies and increased in flux density towards the end of the observation.

\subsection{Stokes $I$ image-plane detections}\label{sec:image_plane}

\begin{figure*}
    \centering
    \includegraphics[width=\textwidth]{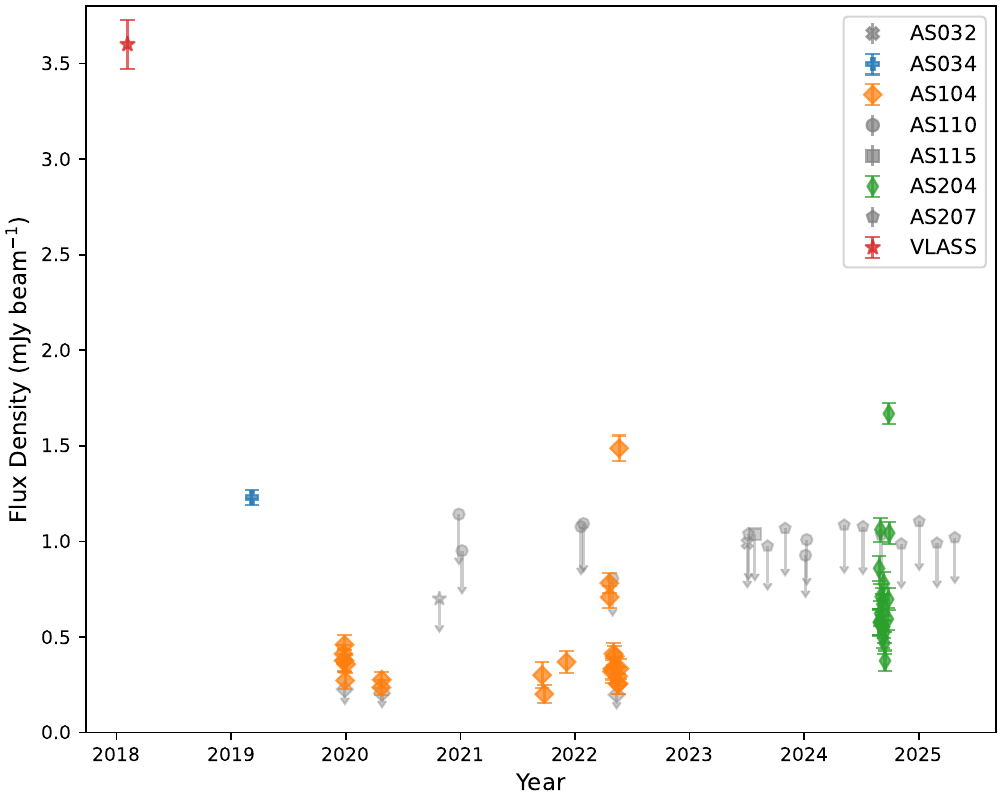}
    \caption{Lightcurve showing Stokes~$I$ image-plane detections (coloured) and non-detections (grey) of \WWPsA\ with ASKAP and the VLA. Uncertainties for the detections were derived by adding the Gaussian 
    fit uncertainty in quadrature with local image RMS noise $\sigma_\text{rms}$. Non-detections are marked as $5\sigma_\text{rms}$ upper limits, with downward arrows of length $\sigma_\text{rms}$. As discussed in \cref{sec:image_plane}, only 22 of the 61 non-detections we found from ASKAP observations are shown.
    The ASKAP observations are labelled by project code: the ASKAP Early Science Broadband Survey (AS034), DINGO (AS104 and AS204), the Rapid ASKAP Continuum Survey \citep[AS110;][]{McConnell2020}, the Survey and Monitoring of ASKAP’s RFI environment and Trends project \citep[AS115;][]{Lourenco2024}, and the Variables and Slow Transients survey \citep[AS032 and AS207;][]{Murphy2021}. 
    }
    \label{fig:WW_PsA_obs_lc}
\end{figure*}

\cref{fig:WW_PsA_obs_lc} shows a lightcurve of detections and non-detections of \WWPsA, made in the Stokes~$I$ image plane with surveys from ASKAP and the \textit{Karl G.\ Jansky} Very Large Array \citep[VLA;][]{Perley2011}. The observations included have a variety of frequency ranges (within 0.8--\SI{4}{GHz}) and durations (\SI{2}{min}--\SI{9}{\hour}). To obtain the ASKAP detections, we cross-matched the proper motion corrected position of \WWPsA\ to the source component catalogue, produced with the source finder package \textsc{selavy} \citep{Whiting&Humphreys2012}, which are publicly available in CASDA. 
Out of 104 ASKAP epochs in which \WWPsA\ was observed, our cross-match yielded 43 detections and 61 non-detections. For the detections, \cref{fig:WW_PsA_obs_lc} shows the peak flux density, with the uncertainty due to the \textsc{selavy} fit added in quadrature with the local root mean square (RMS) noise value, also provided by \textsc{selavy}. For the non-detections we adopted $5\sigma_\text{rms}$ upper limits, where $\sigma_\text{rms}$ is the RMS value of the noiseMap provided by CASDA. To do this, we downloaded $\text{4\arcmin}\times\text{4\arcmin}$ noiseMap cutouts centred on the position of \WWPsA. The RMS of the cutout was therefore the noise local to \WWPsA. For 27 of these non-detections, we were unable to determine $\sigma_\text{rms}$ as the noiseMap was not available on CASDA. 
We excluded a further 12 non-detections with $5\sigma_\text{rms} > \text{\SI{2.9}{\mjpb}}$, as in these observations \WWPsA\ was near the edge of the field, where the image was dominated by artefacts. The remaining 22 ASKAP non-detections are shown in \cref{fig:WW_PsA_obs_lc}.

We performed a similar cross-match to search for detections in the 2--\SI{4}{GHz} VLA Sky Survey \citep[VLASS;][]{Lacy2020} Quick~Look \citep{Gordon2021} catalogues. \WWPsA\ was detected in VLASS Epoch~1 with peak flux density \SI{3.60\pm0.13}{\mjpb}, scaled by a factor of 1.1 as recommended by \citet{Lacy2022}. This detection is reported by \citet{Yiu2024}. We did not obtain a detection in VLASS Epoch~2, with a $5\sigma_\text{rms}$ upper bound of $\sim$\SI{0.7}{\mjpb}.\footnote{For this non-detection, we estimated the local RMS noise at the relevant position and its uncertainty to be the mean and standard deviation, respectively, of the local RMS noise among islands in the relevant Quick~Look subtile (J224641$-$333000), $\sigma_\text{rms} = \SI{137\pm3}{\micro\jansky\per\beam}$.}

\section{Discussion and interpretation}\label{sec:discussion}

\subsection{Significance of volume-limited search}\label{sec:population_statistics}

In the volume-limited search described in \cref{sec:search}, we covered $\text{\SI{2.2e4}{}~star-hours} = \text{\SI{2.5}{}~star-years}$ worth of dynamic spectra and lightcurves. Overall, this was much smaller than the recent \num{100}~star-year search for stellar Type~II radio bursts in Stokes~$V$ dynamic spectra by \citet{Konijn2025}. However, for each star their search covered only $\sim\!\text{\SI{8}{\hour}}$ of observation time, compared to \SI{144}{\hour} in ours. Our search did not account for burst drift rates. 

In our search we made two detections of one star out of our sample of 154 stars. From this, we obtain estimates of the on-sky surface density $n$ and volume density $\rho$ of detected radio stars within \SI{85}{\parsec}:
$\log_{10}(n/\text{deg}^{-2}) = -\text{\num{1.5(7:16)}}$ and $\log_{10}\left(\rho/\text{pc}^{-3}\right) = -\text{\num{3.3(7:16)}}$, with uncertainties given by 95\% Poisson confidence intervals. Based on H$\alpha$ emission, the local density of magnetically active M\;dwarfs within \SI{10}{\parsec} has been estimated to be $\sim\!\text{\SI{0.016}{\per\parsec\tothe{3}}}$ \citep{Villadsen&Hallinan2019,West2011,Henry2018}, implying we detected at most $\sim$20\% of magnetically active M\;dwarfs in our sample volume. However, restricting the sample radius to \SI{25}{\parsec}, which still includes the only detected star, gives a density $\log_{10}\left(\rho/\text{pc}^{-3}\right) = -\text{\num{1.7(7:16)}}$, consistent with up to a 100\% detection rate for magnetically active M\;dwarfs in the volume. This is also consistent with the lower limit of \SI{5e-3}{\per\parsec\tothe{3}} on the density of detectable radio active (0.22--\SI{6}{GHz}, with each star system detected at $\sim\!\text{\SI{1.4}{GHz}}$ in at least one observation) M\;dwarfs within \SI{6.2}{pc} determined by \citet{Villadsen&Hallinan2019}.

\subsection{Properties of \WWPsA}\label{sec:WW_PsA_physical_properties}

The M4V\;dwarf \WWPsA\ is a BY\;Draconis variable \citep{Samus2017} of mass $M_\star = \text{\num{0.19 \pm 0.04}}\:M_\odot$ and radius $R_\star = \text{\num{0.70 \pm 0.13}}\:R_\odot$ \citep{Messina2017} at a distance $d = \text{\SI{20.846(9:10)}{\parsec}}$ \citep{Bailer-Jones2021}. It rotates with period $P_\text{rot} = 2.356 \pm 0.001$\:days \citep{IbanezBustos2023} about an axis at inclination $i = 53^{+22}_{-12}$\,deg to the line of sight \citep{Messina2017}. Radio emission from this star has previously been identified in the Stokes~$I$ image plane \citep{Yiu2024,Driessen2024}, but this emission has not been studied in detail. 
\WWPsA\ forms a binary with the M4.5V~dwarf \TXPsA\ \citep{Song2002}. Both stars are young, sharing a common origin with $\gtrsim\text{40}$ other stars in the $\sim\!\text{\SI{25}{\mega\year}}$-old $\beta$\;Pictoris moving group \citep{Song2003,Torres2006,Shkolnik2012,Messina2017}. The binary is resolved, with a $\sim$\SI{747}{\au} (36\arcsec) separation between the two components \citep{Messina2017}, so the radio emission observed from \WWPsA\ is unlikely to be caused by an interaction between the two components of the system. Owing to its youth and mid-M spectral type, \WWPsA\ shows signatures of enhanced magnetic activity in its chromosphere. It exhibits strong Ca\;\textsc{ii}\;H\&K emission \citep{Beers1996} and X-ray emission \citep{Thomas1998}. 

The measured X-ray flux of \WWPsA\ in the ROSAT All-Sky Survey is $F_\text{X} = \SI{8.21e-12}{\erg\per\second\per\square\centi\metre}$ \citep{Freund2022}, corresponding to a flux $F_\text{X,surf} = \text{\SI{1.4e7}{\erg\per\second\per\square\centi\metre}}$ from the surface of the star. Using the relation \citep{Johnstone&Gudel2015}
\begin{equation}
    \bar{T}_\text{cor} \approx \SI{1.1e5}{} \left(\frac{F_\text{X,surf}}{\SI{}{\erg\per\second\per\square\centi\metre}}\right)^{0.26} \SI{}{\kelvin}
\end{equation}
applicable to \Mdwarfs, we estimate a mean coronal temperature $\bar{T}_\text{cor} \approx \text{\SI{8.0e6}{\kelvin}}$ for \WWPsA. 

\subsection{Coherent and incoherent emission}
\label{sec:SB40848_discussion}

We have shown that \WWPsA\ features both quiescent emission and circularly polarised bursts. In this Section, we separate the quiescent emission from the burst seen in \cref{fig:WW_PsA_flip} and consider the possible emission mechanisms. 

We assumed the quiescent emission to have an effectively constant underlying flux density over the duration of the ASKAP observation SB40848, that the variability in the lightcurve prior to and after the burst was dominated by noise, and that the quiescent emission continued unaffected during the burst. Under these assumptions, we measured the underlying flux density of the quiescent emission in Stokes~$I$ and $V$ by taking the mean flux density across the lightcurves shown in \cref{fig:WW_PsA_flip}, excluding bins between 2022--05--21 20:35 and 23:35 UTC. We obtained $\bar{S}_{I,\text{qui}} = 0.74 \pm 0.03$\,mJy and $\bar{S}_{V,\text{qui}} = -0.02 \pm 0.02$\,mJy, with uncertainties given by the standard error of the mean. The standard deviations among the selected time bins were $\sigma_{I,\text{qui}} = 0.20$\,mJy and $\sigma_{V,\text{qui}} = 0.18$\,mJy, comparable to the root-mean-square noise $\sigma_{I} = 0.22$\,mJy and $\sigma_{V} = 0.15$\,mJy derived from the imaginary components of the visibilities. This justifies our assumption that the variability in the lightcurves, outside the burst's duration, was dominated by noise. We found the quiescent emission to be essentially unpolarised, with $\bar{S}_{V,\text{qui}} / \bar{S}_{I,\text{qui}} = -0.03 \pm 0.03$, compatible with an incoherent emission mechanism.

\begin{figure}
    \centering
    \includegraphics[width=\linewidth]{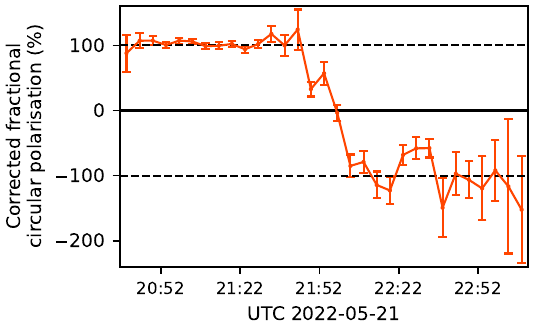}
    \caption{The signed fractional circular polarisation over the duration of the burst in \cref{fig:WW_PsA_flip}, with the flux densities of the quiescent emission subtracted in Stokes~$I$ and $V$, as in \cref{eq:corrected_burst_frac_cp}. The dashed horizontal lines mark the maximum physical absolute fractional polarisation of 100\%.}
    \label{fig:burst_frac_cp}
\end{figure}

We analysed the fractional circular polarisation during the burst with the quiescent emission subtracted, computing
\begin{equation}\label{eq:corrected_burst_frac_cp}
    \frac{S_V - \bar{S}_{V,\text{qui}}}{S_I - \bar{S}_{I,\text{qui}}}
\end{equation}
for each time bin during the burst. The result is shown in \cref{fig:burst_frac_cp}. Both before and after the inversion, the burst had $\sim$100\% fractional circular polarisation, with a tighter constraint prior to the inversion due to the higher flux density. The bursts in \cref{fig:WW_PsA_I_V_ds_and_lc_SB65405_SB66106_tav30} were also $\sim$100\% circularly polarised. Such high fractional circular polarisation requires a coherent emission mechanism for the bursts.

\subsection{Quiescent emission}

Quiescent emission from \Mdwarfs\ is understood to originate from gyrosynchrotron radiation of mildly relativistic electrons with a power-law energy distribution \citep{Kundu&Shevgaonkar1985} of spectral index $\delta \approx 3$ \citep{White1989}. Assuming a magnetic field dependence $B \propto r^{-n}$, where $r$ is the distance from a buried dipole at depth $r_0$, and a collisionless plasma with density $N \propto r^{-m}$, \citet{White1989} find the peak frequency of the quiescent emission to be
\begin{equation}\label{eq:qui_peak_freq}
    \small
    \frac{\nu_\text{peak}}{\SI{}{\giga\hertz}} = 
    \left[
    \frac{10^{2.72+0.33\delta}
    \frac{N_0}{\SI{1e5}{\per\centi\metre\tothe{3}}}
    \frac{r_0}{\SI{1e9}{\centi\metre}}
    }
    {m+n(0.30+0.98\delta)-1.00}
    \left( \frac{B_0}{\SI{}{\kilo\gauss}} \right)^{0.30+0.98\delta}
    \right]^\frac{1}{1.30+0.98\delta}.
\end{equation}
Here, $N_0$ is the density of electrons with energies above a low-energy cutoff at the surface of the star and $B_0$ is the equatorial surface magnetic field. 

Our ATCA observation showed the quiescent emission from \WWPsA\ increased with frequency, implying high optical depth over the observed band, with $\nu_\text{peak} \gtrsim \SI{3}{\giga\hertz}$. In the case of a global dipolar magnetic field, i.e. $r_0 = R_\star = \SI{0.70\pm0.13}{R_\odot}$ \citep{Messina2017} and $n=3$, taking typical values of the remaining parameters ($m=3$, $\delta=3$, $N_0 = \SI{1e5}{\per\centi\metre\tothe{3}}$) gives a loose lower bound on the field of $B_0 \gtrsim \SI{200}{\gauss}$. Note that in \cref{eq:qui_peak_freq}, $\nu_\text{peak}$ is much more strongly dependent on $B_0$ than the other variables, with the exception of $\delta$. For example, increasing $N_0$ by a factor of 10 only reduces the lower bound on $B_0$ by a factor of 2. Setting $\delta = 2$ decreases the lower bound to $\sim\!\SI{70}{\gauss}$. However, \citet{White1989} find that models with $\delta = 2$ give flux densities much greater than those observed from real stars. Indeed, scaling the model flux densities in Tables~2, 3, and 5 of \citet{White1989} by a factor $(R_\star/\SI{1e10}{\centi\metre})^{2.46} (d/\SI{2.7}{\parsec})^{-2}\approx0.8$ (using $S \sim r_0^{2.46}$ for optically thick emission with $\delta=2, m=n=3$ from Table~1 in the same work), the resulting flux densities at \SI{1.5}{\giga\hertz} are much greater than that observed at this frequency in \cref{fig:ATCA_qui}. This confirms that $\delta \approx 3$ is a more realistic scenario for the quiescent emission from \WWPsA. The bound $B_0 \gtrsim \SI{200}{\gauss}$ is unsurprising, as \SI{}{\kilo\gauss}-scale average magnetic fields strengths have been measured from many \Mdwarfs\ \citep[e.g.][]{Kochukhov2021}.

Among the observations from ASKAP in \cref{fig:WW_PsA_obs_lc}, \WWPsA\ was only detected in those with long durations (8--\SI{9}{\hour}). We posit this to be due to quiescent emission below the $\sim$\SI{1}{\mjpb} detection thresholds for the observations with shorter durations. Note, however, that while \cref{fig:WW_PsA_flip} and our follow-up observation confirm the presence of quiescent emission, we cannot exclude the possibility of these detections being caused by unpolarised bursts, except in the observations in which we produced Stokes~$I$ dynamic spectra (see \cref{tab:WWPsAobservations}).

\subsection{Coherent bursts}

There are two mechanisms capable of producing the highly polarised coherent radio bursts we have observed from \WWPsA: ECM emission and plasma emission. In this section, we calculate the magnetic field and plasma density required at the emission region in each scenario and discuss \changed{physical models which can explain the observed properties of the bursts.}

ECM emission produces narrow-band emission close to harmonics of the local electron cyclotron frequency,\footnote{Here we use the Gaussian system for electromagnetic quantities.}
\begin{equation}
    \nu_\text{c} = \frac{eB}{2\pi m_\text{e} c} 
\end{equation}
where $e$ and $m_\text{e}$ are the charge and mass of the electron, respectively, $c$ is the speed of light\changed{, and $B$ is the local magnetic field strength at the emission site}. Amplification is only expected to be significant for the first and second harmonics \citep{Dulk1985}. The maser growth rate is greatest at the first harmonic, but the second harmonic is considered more likely to be observed due to strong gyroresonance absorption of the fundamental \citep{Melrose&Dulk1982}. If this is the emission mechanism, the observed frequency range of 891--\SI{1137}{\mega\hertz} for the burst in \cref{fig:WW_PsA_flip} implies a range of magnetic field strengths over the emission region of 320--\SI{410}{\gauss} at the fundamental, or 160--\SI{200}{\gauss} for emission at the second harmonic. In \cref{fig:WW_PsA_I_V_ds_and_lc_SB65405_SB66106_tav30}, the emission covers 1315--\SI{1439}{\mega\hertz}, requiring magnetic field strengths of 470--\SI{510}{\gauss} at the fundamental or 230--\SI{260}{\gauss} at the second harmonic.

ECM emission is most efficient when $\nu_\text{p} \ll \nu_\text{c}$ \citep{Melrose1984}, where
\begin{equation}\label{eq:plasma_freq}
    \nu_\text{p} = \sqrt{\frac{n_\text{e}e^2}{\pi m_\text{e}}}
\end{equation}
is the electron plasma frequency and $n_\text{e}$ is the total number density of the electrons. For the radiation at $\nu = \SI{891}{\mega\hertz}$ in \cref{fig:WW_PsA_flip}, this would imply $n_e \ll \SI{9.8e9}{\per\cubic\centi\metre}$ if $\nu \simeq \nu_c$, or $n_e \ll \SI{2.5e9}{\per\cubic\centi\metre}$ if $\nu \simeq 2\nu_c$.

Alternatively, if the observed bursts are due to plasma emission, their high fractional polarisation of $\sim$100\% can only be explained by emission at the fundamental of the plasma frequency \citep{Dulk1985}. Thus, from \cref{eq:plasma_freq}, the burst in \cref{fig:WW_PsA_flip} would correspond to a range 0.98--\SI{1.6e10}{\per\cubic\centi\metre} of plasma densities over the emission region, while those in \cref{fig:WW_PsA_I_V_ds_and_lc_SB65405_SB66106_tav30} would correspond to 2.2--\SI{2.6e10}{\per\cubic\centi\metre}. These are in agreement with typical stellar coronal densities as measured from the X-ray \ion{O}{vii} and \ion{Ne}{ix} lines \citep{Ness2004}.

\changed{
The rate of detectable radio bursts from \WWPsA\ is very low. The observations listed in \cref{tab:WWPsAobservations} include \SI{195.3}{\hour} covering the band 1296.5--\SI{1440.5}{\mega\hertz}, in which we detected two long-duration \citep[$\gtrsim\text{\SI{30}{\minute}}$, adopting the terminology of][]{Villadsen&Hallinan2019} circularly polarised bursts. Treating the bursts as stochastic for now and using a 95\% Poisson confidence interval, we estimate they occur at a rate of \SI{0.25(64:22)}{\per\day}. That is, we expect there to be one burst detectable in Stokes~$V$ for every \SI{100(710:70)}{\hour} of observation time in this band.}

\changed{
For comparison, \citet{Villadsen&Hallinan2019} conducted observations of the M3V dwarf \ADLeo, with \SI{22}{\hour} covering the 1296.5--\SI{1440.5}{\mega\hertz} band. They detected three long-duration bursts in the relevant frequency range, each with maximum $S_V\gtrsim\text{\SI{5}{\milli\jansky}}$, well above our typical detection threshold. This yields one burst for every \SI{7.3(282:48)}{\hour} of observation time. A similar treatment for the observations of the M5.5V dwarf \UVCet\ in the same work implies one burst every \SI{2.4(29:12)}{\hour} (with a long-duration burst every \SI{3.7(63:20)}{\hour}). The higher measured burst rates of these stars compared to \WWPsA\ can in part be attributed to their distances \citep[\SI{4.9651(7)}{\parsec} and \SI{2.675(4)}{\parsec}, respectively;][]{GaiaCollaboration2023}, as not all of the events reported by \citet{Villadsen&Hallinan2019} would remain above our detection threshold when scaled to the distance of \WWPsA. It may also be contributed to by physical differences such as the rotation period, which is much shorter for \UVCet, at \SI{0.2268(3)}{\day} \citep{Barnes2017}, than for \WWPsA.
}

\changed{
The consistent absolute fractional polarisation of $\sim$100\% for all three bursts from \WWPsA\ seen in \cref{fig:WW_PsA_I_V_ds_and_lc_SB65405_SB66106_tav30,fig:burst_frac_cp} is unusual behaviour for a magnetically active M\;dwarf. The coherent bursts from \UVCet, \ADLeo\ \citep{Villadsen&Hallinan2019}, EV\;Lac \citep{White&Jackson&Kundu1989,Osten2005,Lu2025}, and Proxima Centauri \citep{Perez-Torres2021,Zic2020} mostly feature fractional polarisations below $\sim$85\%, and are consistent with 100\% in only a small number of cases. However, in some cases this may be due to underlying quiescent emission, which we subtracted to produce \cref{fig:burst_frac_cp}. All of these systems also produce short-duration bursts, which we did not detect from \WWPsA.}

Long-duration coherent radio bursts from \Mdwarfs\ have consistently been found to be due to ECM emission \citep{Lynch2017,Villadsen&Hallinan2019,Bloot2024}, so there is a strong possibility this is also the case for \WWPsA. Evidence of ECM emission, rather than plasma emission, being the mechanism for a coherent burst can be found in a number of ways. These include, with varying levels of stringency:
\begin{enumerate}
    \item identifying fine temporal structure (\SI{}{\milli\second}-timescale) in the dynamic spectrum \citep[e.g.][]{Osten&Bastian2008};
    \item combining the sense of the circular polarisation with information about the star's magnetic field structure \citep[e.g. from Zeeman~Doppler imaging;][]{Morin2008,Kochukhov&Lavail2017,Villadsen&Hallinan2019} to determine if it is in the sense of the \textit{x}-mode or the \textit{o}-mode, which are favoured \changed{under typical conditions} for ECM and plasma emission, respectively \changed{\citep{Dulk1985}};
    \item observing elliptical polarisation, which indicates the emission to be in the sense of the \textit{x}-mode \citep{Melrose&Dulk1991,Zic2019}; and
    \item confirming the bursts \changed{are emitted in a highly beamed manner, particularly by identifying rotational modulation}. Although plasma emission can be beamed tangentially to the magnetic field, as occurs in the case of some solar Type~III radio bursts \citep{Thejappa2012}, \changed{narrow angular beaming is intrinsic to the ECM mechanism, making it the favourable explanation, especially when the emission is} beamed at large angles to the magnetic field \citep[e.g.][]{Melrose&Dulk1982}.
\end{enumerate}
Given the available ASKAP observations have a time resolution of \SI{10}{s}, that Zeeman~Doppler imaging is unavailable, and because we did not detect linear polarisation, we are left with only rotational modulation as an avenue for finding direct evidence that the bursts from \WWPsA\ are due to ECM emission. We investigate this possibility in the following sections. 

\subsubsection{Polarity reversal}\label{sec:polarity_reversal}

Polarity reversal such as that shown in \cref{fig:WW_PsA_flip} has been observed previously in rotationally modulated ECM emission at \SI{154}{\mega\hertz} from \UVCet\ \citep{Lynch2017}. At \SI{888}{\mega\hertz}, elliptically polarised bursts have been observed from \UVCet, indicating an auroral cavity with density more than six orders of magnitude lower than would otherwise be expected at the source altitude \citep{Zic2019}. 

An auroral oblique rotator model \citep[e.g.][]{Leto2016} is therefore appropriate for \UVCet. Under this model, ECM emission is produced from auroral rings within cavities above each magnetic pole of the star, emitted in planes roughly perpendicular to the magnetic axis. Thus, a pair of pulses with opposite circular polarisation is observed near each magnetic null, that is, near the rotational phases at which the average line-of-sight stellar magnetic field is zero. The pulses may then be refracted in the stellar magnetosphere, increasing the delay between their arrival times. If the pulses are sufficiently separated, this results in an observed burst with polarity reversal at each magnetic null.

The strong dipolar magnetic field of \UVCet\ \citep{Kochukhov&Lavail2017} is consistent with this model. Rotationally modulated auroral ECM radiation of this type is well understood to be emitted by Earth, Jupiter, Saturn, Uranus, and Neptune \citep[for a review, see][]{Zarka1998}. It has also been observed from some high-mass magnetic chemically peculiar (MCP) stars \citep[e.g.][]{Trigilio2011,Das2024}. 

At frequencies above \SI{744}{\mega\hertz}, the bursts from \UVCet\ have circular polarisation which is consistently right-handed \citep{Villadsen&Hallinan2019,Zic2019}, whereas \citet{Lynch2017} observed polarity reversals at 139--\SI{169}{\mega\hertz}. 
This behaviour is similar to our observation of polarity reversal in the bursts from \WWPsA\ at lower frequencies (891--\SI{1137}{\mega\hertz}), but not at higher frequencies (1315--\SI{1439}{\mega\hertz}). Likewise, in the case of the MCP stars HD\;133880 and \CUVir, the oppositely polarised pulses are observed up to different maximum frequencies \citep{Das2020,Das&Chandra2021}. This may be due to asymmetry in the magnetic field or the magnetosphere \citep{Das&Chandra2021}.

Polarity reversal has also been observed in bursts of rotationally modulated ECM emission from the ultracool dwarfs TVLM\;513$-$46546 \citep{Hallinan2007,Lynch2015} and 2MASS\;J0746425$+$200032 \citep{Lynch2015}. In both cases, \citet{Lynch2015} found that the observed dynamic spectra are much better explained by a sequence of three distinct emission sites, corresponding to separate coronal loops. Thus, \changed{rotational modulation} does not necessarily require the emission to be auroral in nature. This means that although auroral emission is perhaps the simplest \changed{geometric model that can explain} the observed polarity reversal from \WWPsA, it is not the only plausible explanation, even in the case of rotationally modulated bursts of ECM emission. \changed{However, any geometric model for the polarity reversal must rely on narrow angular beaming, which is more easily explained by ECM emission than plasma emission.}

\changed{Alternatively, the polarity reversal may be caused by a change in the dominant magneto-ionic mode emitted due to variations in the properties of the plasma in the emission region. For example, consider the case of emission from a cyclotron maser driven by a loss cone. The growth rates of the modes are strongly dependent on the ratio $\nu_\text{p}/\nu_\text{c}$, with the fundamental \textit{x}-mode becoming suppressed for $\nu_\text{p}/\nu_\text{c} \gtrsim 0.3$ \citep{Melrose1984}. A rapid change in $n_\text{e}$ or $B$ could cause the dominant mode to change from the \textit{x}-mode to the \textit{o}-mode, or vice versa, resulting in the polarity reversal seen in \cref{fig:WW_PsA_flip}. Significant ECM emission in the sense of the \textit{o}-mode is known to be possible under stellar conditions, as it has been observed from the early-type star HD\;142301 \citep{Leto2019}. With the parameters considered by \citet{Sharma1984}, for a saturated maser the predicted polarisation fraction transitions sharply from 100\% in the sense of the \textit{x}-mode to 100\% in the sense of the \textit{o}-mode at $\nu_\text{p}/\nu_\text{c} = 0.35$. Under such conditions, a transition from, say, $\nu_\text{p}/\nu_\text{c} = 0.33$ to $0.37$ (e.g.\ due to a $\sim$25\% increase in $n_\text{e}$ or a $\sim$10\% decrease in $B$) over the course of $\sim$\SI{20}{\minute} could therefore explain the behaviour of the fractional polarisation shown in \cref{fig:burst_frac_cp}. The power emitted in the \textit{x}-mode when it is dominant is expected to be greater than that in the \textit{o}-mode when it is dominant \citep{Sharma1984}. Thus, in this scenario, we suggest that the higher flux density prior to the polarity reversal (see \cref{fig:WW_PsA_flip}) is in better agreement with a transition from the \textit{x}-mode to the \textit{o}-mode than vice versa, i.e.\ an increase in $n_\text{e}$ or a decrease in $B$.}

\changed{
However, in the case of plasma emission, such a transition in the dominant magneto-ionic mode cannot explain the behaviour of the fractional polarisation in \cref{fig:burst_frac_cp}. This is because fundamental plasma emission occurs below the right-hand cutoff of the \textit{x}-mode \citep{Dulk1985,Chen2016},
\begin{equation}
    \nu_\text{R} = \frac{1}{2} \left( \nu_\text{c} + \sqrt{\nu_\text{c}^2 + 4\nu_\text{p}^2} \right) > \nu_\text{p},
\end{equation}
so the emission can only be in the sense of the \textit{o}-mode,\footnote{\changed{The \textit{z}-mode, which is the lower branch of the \textit{x}-mode, can propagate at $\nu = \nu_\text{p}$ \citep{Benz2002,Chen2016}. However, it cannot directly escape the plasma \citep[e.g.][]{Melrose1984,Treumann2006}.}} while at the second harmonic the radiation is not expected to have such high fractional polarisation \citep[e.g.][]{Melrose1978,Melrose1980,Dulk1985}.
}

\subsubsection{Rotational modulation}\label{sec:rotational_modulation}

We investigated the possibility that the observed bursts from \WWPsA\ were rotationally modulated. 
\changed{We used} the photometric rotation period $P_\text{rot} = \SI{2.356\pm0.001}{\day}$ determined by \citet{IbanezBustos2023} \changed{with data from the Transiting Exoplanet Survey Satellite \citep[TESS;][]{Ricker2015}. However, we note that the uncertainty in this measurement does not account for the effect of differential rotation of starspots \citep[e.g.][]{CollierCameron2002}, so the results following from it are tentative. For} sufficient $N \in \mathbb{Z}$ we calculated the Barycentric Julian Dates (BJDs)
\begin{equation}\label{eq:ephemeris}
    \text{BJD}_N = \text{BJD}_0 + NP_\text{rot},
\end{equation}
corresponding to the same rotational phase $\phi_0$ as the centre of the burst in \cref{fig:WW_PsA_flip}, at time $\text{BJD}_0 = \text{\SI[uncertainty-mode=compact]{2459721.410(7)}{\day}}$. We compared these to the BJDs of the bursts in \cref{fig:WW_PsA_I_V_ds_and_lc_SB65405_SB66106_tav30}: \SI{2460554.00}{}\,--\,\SI{2460554.07}{\day} for the earlier burst and \SI{2460581.28}{}\,--\,\SI{2460581.31}{\day} for the later burst. Doing so, we found $\text{BJD}_{365} = \text{\SI[uncertainty-mode=compact]{2460581.35(37)}{\day}}$ to be consistent with the time of the later burst, whereas the earlier burst was closest to $\text{BJD}_{353} = \text{\SI[uncertainty-mode=compact]{2460553.08(35)}{\day}}$, offset from its centre by $2.7\sigma_{\text{BJD}_{353}}$. The later burst can therefore plausibly be explained by rotationally modulated pulses from the same emission region as that for the burst in \cref{fig:WW_PsA_flip}, but this is unlikely for the earlier burst.

If the emission is rotationally modulated, it is possible to place further constraints on the properties of the star. We denote our updated quantities with a prime. Enforcing $\text{BJD}_{365}' = \text{BJD}_\text{later} = \text{\SI[uncertainty-mode=compact]{2460581.30(2)}{\day}}$, where $\text{BJD}_\text{later}$ is the BJD of the later burst in \cref{fig:WW_PsA_I_V_ds_and_lc_SB65405_SB66106_tav30} (covering the possibility that its end corresponds to the end of the centre of the burst in \cref{fig:WW_PsA_flip}), we consider a new ephemeris,
\begin{equation}\label{eq:updated_ephemeris}
    \text{BJD}_N' = \text{BJD}_0 + NP_\text{rot}',
\end{equation}
to obtain $P_\text{rot}' = \text{\SI[uncertainty-mode=compact]{2.35586(6)}{\day}}$. \changed{We work with this rotation period moving forward, but caution that it is both model-dependent and reliant on the optical period reported by \citet{IbanezBustos2023}.} Next, we adopt the projected rotational velocity $v \sin i = \text{\SI{12.0(1.1)}{\kilo\metre\per\second}}$, which \citet{Messina2017} infer from 
\changed{prior work} \citep{Christian&Mathioudakis2002,Torres2006,Jayawardhana2006,Bailey2012}, where $i$ is the inclination of the rotation axis to the line of sight. Then, 
\begin{equation}
    \sin i' = 
    \frac{P_\text{rot}'}{2 \pi R_\star}
    \left( v \sin i \right)
\end{equation}
gives $\sin i' = \text{0.80} \pm \text{0.17}$, or $i' = \text{53}^{+\text{22}}_{-\text{14}}$\,\SI{}{\deg}. This does not improve the precision of the measured inclination compared to that found by \citet{Messina2017}, as the dominant sources of uncertainty are $v \sin i$ and $R_\star$.

The burst on 2025--09--13 (see \cref{sec:archival_observations}) had a BJD range spanning from \SI[uncertainty-mode=compact]{2460932.19}{\day} to at least \SI[uncertainty-mode=compact]{2460932.39}{\day}. This matches $\text{BJD}_{514}' = \text{\SI[uncertainty-mode=compact]{2460932.32(0.03)}{\day}}$ from \cref{eq:updated_ephemeris}. Using an alternative ephemeris given by 
\begin{equation}
    \text{BJD}_M = \text{BJD}_\text{later} + MP_\text{rot},
\end{equation}
where $M \in \mathbb{Z}$, yields $\text{BJD}_{M=149} = \text{\SI[uncertainty-mode=compact]{2460932.34(15)}{\day}}$. With either method, the arrival time of the burst on 2025--09--13 is successfully predicted. We therefore \changed{tentatively claim that the bursts are} composed of rotationally modulated pulses. This is most easily explained by the intrinsic narrow angular beaming of ECM emission.

The earlier burst in \cref{fig:WW_PsA_I_V_ds_and_lc_SB65405_SB66106_tav30}, at time $\text{BJD}_\text{earlier} = \text{\SI[uncertainty-mode=compact]{2460554.04(4)}{\day}}$, differs in phase from the later burst (and thus also the burst in \cref{fig:WW_PsA_flip}) according to
\begin{equation}
    \Delta\phi = \left\{ \frac{\text{BJD}_\text{earlier} - \text{BJD}_\text{later}}{P_\text{rot}'} \right\},
\end{equation}
where $\{x\} = x- \lfloor x \rfloor$ denotes the fractional part of $x$. We obtain $\Delta\phi = \text{0.429} \pm \text{0.019}$.

Among the observations listed in \cref{tab:WWPsAobservations}, the ephemeris given by \cref{eq:updated_ephemeris} predicts bursts in the ASKAP scheduling blocks SB65347 for $N=353$, SB65665 for $N=356$, and SB65818 for $N=359$. Additionally, if the earlier burst in \cref{fig:WW_PsA_I_V_ds_and_lc_SB65405_SB66106_tav30} is also rotationally modulated, we expect bursts at the times $\text{BJD}_0 + (N+\Delta\phi)P_\text{rot}'$. We therefore also predict bursts in SB40196 for $N=-8$, SB40880 for $N=0$, and SB65836 for $N=359$. However, we did not detect polarised emission in any of these observations. \changed{If the emission is rotationally modulated, this} may be due to intrinsic variability between pulses caused by instability in the emission region, which occurs, for example, in the case of the ECM pulses from \CUVir\ \citep[e.g.][]{Ravi2010}. \citet{Trigilio2011} note that the ECM growth rate is a nonlinear function of various parameters \changed{\citep[e.g.][]{Melrose1984,Sharma1984,Mutel2007}}, including $n_\text{e}$ and \changed{$B$}, as well as the electron velocity distribution and its anisotropy. Small variations in these parameters can therefore cause the large variations in the observed flux density. 
\changed{In addition, ECM emission requires acceleration or heating of electrons, for example resulting from magnetic field reconnection \citep{Dulk1985}. It may be that the observed bursts were enabled by such events. Magnetic reconnection is also responsible for stellar flares \citep[e.g.][]{Parker1963}, several of which have been identified from \WWPsA\ \citep{IbanezBustos2023,Lin2024} in observations from TESS.} 

\subsubsection{Oblique rotator model}

For an oblique rotator model \citep[e.g.][]{Leto2016}, in which the bursts are auroral and the star has a dipolar field, there is a magnetic null at $\phi_0$. Then, taking the earlier burst in \cref{fig:WW_PsA_I_V_ds_and_lc_SB65405_SB66106_tav30} to be at the other null and choosing the phases such that it occurs at $1-\phi_0$ gives $\phi_0 = (1-\Delta\phi)/2 = \text{0.286} \pm \text{0.009}$. 
Radiation emitted at an angle $\theta_\text{M}$ to the magnetic dipole axis is observed at phase $\phi$, where \citep{Trigilio2000,Leto2016}
\begin{equation}\label{eq:cos_theta_M}
    \cos\theta_\text{M} = \sin\beta \sin i' \cos 2\pi\phi + \cos\beta \cos i'.
\end{equation}
The obliquity $\beta$ is the angle between the rotation axis and the magnetic dipole axis. The magnetic nulls correspond to $\cos\theta_\text{M} = 0$, implying \citep{Leto2016}
\begin{equation}\label{eq:obliquity}
    \beta = \arctan\left(\frac{-1}{\tan i' \cos2\pi\phi_0}\right).
\end{equation}
This yields $\tan\beta = \text{3.4} \pm \text{2.1}$, or $\beta = \text{74}^{+\text{6}}_{-\text{22}}\,\text{\SI{}{\deg}}$. Such a high obliquity would be unusual for a star of type $\sim\!\text{M4}$. Among those investigated by \citet{Morin2008}, most featured fields with strong dipolar components closely aligned with the rotation axis (i.e.\ low $\beta$), with the exception of EV\;Lac, which had the longest rotation period of \SI{4.38}{\day} (compared to \num{0.40}--\SI{2.77}{\day} for the other five stars). We note that these stars have masses in the range 0.25--0.42\,$M_\odot$, higher than that of \WWPsA\ at $\num{0.19(4)}\,M_\odot$ \citep{Messina2017}.

We can now determine the range of angles $\theta_\text{M}$ which the burst in \cref{fig:WW_PsA_flip} corresponds to under the oblique rotator model. The right circularly polarised (RCP) and left circularly polarised (LCP) pulses lasted $\sim\!\text{\SI{1.3}{\hour}}$ prior to and after the null at $\phi_0$, respectively. Substituting the corresponding range of phases into \cref{eq:cos_theta_M}, we find that the start of the RCP pulse corresponds to $\theta_\text{M} = \text{\SI{83.8(3.8)}{\deg}}$, while the end of the LCP pulse corresponds to $\theta_\text{M} = \text{\SI{96.0(3.8)}{\deg}}$. That is, the maximum upward deflection of the radiation from the plane perpendicular to the dipole axis was \SI{6.2(3.8)}{\deg} for the RCP pulse and \SI{6.0(3.8)}{\deg} for the LCP pulse.

\subsubsection{Pulse spectral properties}

\begin{figure}
    \centering
    \includegraphics[width=\linewidth]{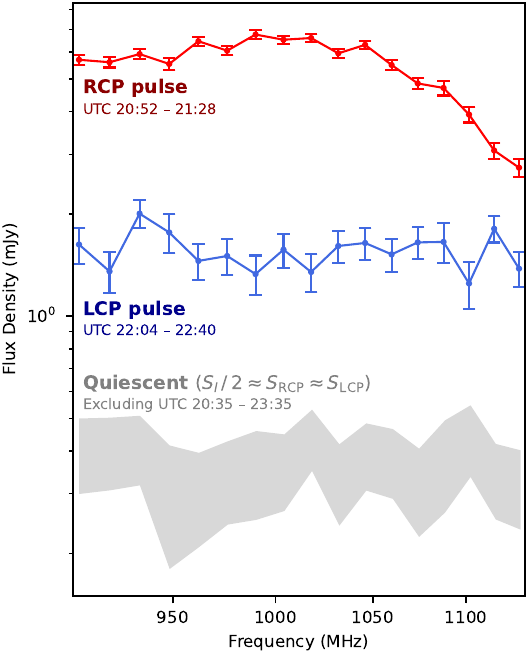}
    \caption{Spectra of $S_\text{RCP}$ and $S_\text{LCP}$ at the peaks of the RCP and LCP pulses, respectively, for the burst shown in \cref{fig:WW_PsA_flip}. Also shown is the spectrum of the quiescent emission, for which we use $S_I / 2$ to estimate both $S_\text{RCP}$ and $S_\text{LCP}$. The grey band indicates the range $S_I / 2 \pm \sqrt{\sigma_{S_I/2}^2 + \sigma_{S_V/2}^2}$. All times shown are for the date 2022--05--21~UTC. For the quiescent emission, the whole observation was included except the period indicated.}
    \label{fig:SB40848_RCP_LCP_spec}
\end{figure}

For the dynamic spectra in \cref{fig:WW_PsA_flip}, we performed a linear transformation from the Stokes~$I$ and $V$ flux densities to those of the RCP and LCP components of the field,\footnote{This normalisation differs by a factor of two from that used in some works \citep[e.g.][]{Villadsen&Hallinan2019,Das2025}, but is in agreement with others \citep[e.g.][]{Leto2016,Myserlis2018}. Under our definition, $S_\text{RCP}$ and $S_\text{LCP}$ measure the contributions to the flux density from photons of left-handed and right-handed helicity, respectively.}
\begin{subequations}\label{eq:IandVtoRCPandLCP}
\begin{equation}
    S_\text{RCP} = \frac{S_I + S_V}{2}
\end{equation}
and 
\begin{equation}
    S_\text{LCP} = \frac{S_I - S_V}{2}.
\end{equation}
\end{subequations}
This allowed us to separate the RCP and LCP pulses. At the peak of each pulse (2022--05--21~20:52 to 21:28~UTC for the RCP pulse and 22:04 to 22:40~UTC for the LCP pulse) we created a spectrum, shown in \cref{fig:SB40848_RCP_LCP_spec}. Also shown in \cref{fig:SB40848_RCP_LCP_spec} is the spectrum of the quiescent emission in the same observation, where we have used $S_I/2$ to estimate $S_\text{RCP}$ and $S_\text{LCP}$ given its low fractional polarisation, as discussed in \cref{sec:SB40848_discussion}. 

The quiescent emission did not show significant variation over the observed bandwidth. The RCP pulse behaved similarly to the auroral pulses emitted by some MCP stars \citep[e.g.][]{Das&Chandra2021}, featuring a roughly constant flux density followed by a decrease beginning at $\sim\!\text{\SI{1050}{\mega\hertz}}$. By fitting a power law for the latter part of the spectrum, we estimated the RCP pulse should merge with the $\sim\!\text{\SI{0.4}{\milli\jansky}}$ level of the quiescent emission at 1300--\SI{1400}{\mega\hertz}. The LCP pulse did not exhibit the same behaviour, remaining approximately constant over the band.

\section{Summary}\label{sec:conclusion}

In this work we have identified coherent radio bursts and quiescent emission from \WWPsA. This resulted from a \SI{2.2e4}{}~star-hour volume-limited search for polarised stellar radio emission in the ASKAP DINGO survey. For \WWPsA, we searched for polarised bursts in the dynamic spectra and lightcurves from \SI{283.3}{\hour} of observations. We detected three bursts in this search, with a fourth identified in a later DINGO observation with the VASTER pipeline. 

The quiescent emission varied slowly on a timescale of hours. As we found its peak frequency to be $\gtrsim \text{\SI{3}{\giga\hertz}}$, we expect the surface magnetic field at the relevant sites of emission to be in the hundreds of gauss. 

One of the bursts exhibited polarity reversal, a common feature of rotationally modulated ECM emission. We found the bursts to be consistent with rotationally modulated emission \changed{given the photometric rotation period reported by \citet{IbanezBustos2023}. Using the radio bursts, we derived an updated} rotation period $P_\text{rot}' = \text{\SI{2.35586(6)}{\day}}$ \changed{and} inclination $i' = \text{\SI{53(22:14)}{\deg}}$. Under an oblique rotator model, we \changed{also} derived the obliquity $\beta = \text{\SI{74(6:22)}{\deg}}$ and the maximum upward deflection angles of \SI{6.2(3.8)}{\deg} and \SI{6.0(3.8)}{\deg} for the RCP and LCP pulses, respectively. The \changed{required} narrow angular beaming suggests ECM emission as the most likely mechanism \changed{if the bursts are rotationally modulated}. For second harmonic ECM emission, the observed frequency ranges require field strengths at the emission sites of 160--\SI{200}{\gauss} and 230--\SI{260}{\gauss}. Our ephemeris predicts burst arrival times during some observations in which we made no detection, \changed{which may be due to a variable maser growth rate caused by changing conditions in the plasma, also seen in} the auroral ECM pulses from MCP stars. One of the observed pulses from \WWPsA\ also shows similar spectral behaviour to pulses from MCP stars. 

\changed{Under an alternative model in which the polarity reversal is due to a change in the dominant magneto-ionic mode of the emission, we still favour the ECM mechanism. This is because plasma emission cannot explain a polarity reversal from 100\% fractional polarisation in the sense of one mode to 100\% in the sense of the opposite mode.}

\section*{Acknowledgements}
\changed{We thank the reviewer for their suggestions regarding our discussion of physical models for the bursts from \WWPsA.}

\changed{JM-B is supported by an Australian Government Research Training Program (RTP) Scholarship.}

Parts of this research were supported by the Australian Research Council Centre of Excellence for Gravitational Wave Discovery (OzGrav), project number CE230100016. 

This scientific work uses data obtained from Inyarrimanha Ilgari Bundara, the CSIRO Murchison Radio-astronomy Observatory. We acknowledge the Wajarri Yamaji People as the Traditional Owners and native title holders of the Observatory site. CSIRO’s ASKAP radio telescope is part of the Australia Telescope National Facility (\href{https://ror.org/05qajvd42}{https://ror.org/05qajvd42}). Operation of ASKAP is funded by the Australian Government with support from the National Collaborative Research Infrastructure Strategy. ASKAP uses the resources of the Pawsey Supercomputing Research Centre. Establishment of ASKAP, Inyarrimanha Ilgari Bundara, the CSIRO Murchison Radio-astronomy Observatory and the Pawsey Supercomputing Research Centre are initiatives of the Australian Government, with support from the Government of Western Australia and the Science and Industry Endowment Fund.

The Australia Telescope Compact Array is part of the Australia Telescope National Facility (\href{https://ror.org/05qajvd42}{https://ror.org/05qajvd42}) which is funded by the Australian Government for operation as a National Facility managed by CSIRO. We acknowledge the Gomeroi people as the Traditional Owners of the Observatory site.

This paper includes archived data obtained through the CSIRO ASKAP Science Data Archive, CASDA (\href{http://data.csiro.au}{http://data.csiro.au}).

The National Radio Astronomy Observatory and Green Bank Observatory are facilities of the U.S. National Science Foundation operated under cooperative agreement by Associated Universities, Inc.

\section*{Data Availability}
The ASKAP data used in this paper are public and can be accessed through the CSIRO
ASKAP Science Data Archive
(CASDA\footnote{\url{https://data.csiro.au/domain/casdaObservation}})
under project codes AS032, AS034, AS104, AS110, AS115, AS204, and AS207. 
The ATCA data and subsequent analysis are available upon request. The VLASS Quick Look catalogues used are publicly available from the Canadian Initiative for Radio Astronomy Data Analysis (CIRADA) website\footnote{\url{https://cirada.ca/vlasscatalogueql0}}.

\bibliographystyle{pasa-mnras}
\bibliography{bibliography}

\end{document}